\documentclass[twocolumn,showpacs,preprintnumbers,amsmath,amssymb]{revtex4}

\usepackage{graphicx}
\usepackage{dcolumn}
\usepackage{bm}

\begin{document}

\title{Influence functional approach to decoherence during Inflation}

\author{Fernando C. Lombardo}
\email{lombardo@df.uba.ar}

\affiliation{Departamento de F\'\i sica {\it Juan Jos\'e Giambiagi}, FCEyN UBA,
Facultad de Ciencias Exactas y Naturales, Ciudad Universitaria,
Pabell\' on I, 1428 Buenos Aires, Argentina}

\date{\today}

\begin{abstract}
We show how the quantum to classical transition of the cosmological 
fluctuations produced during inflation can be described 
by means of the influence functional and the master equation. We split the inflaton 
field into the system-field 
(long-wavelength modes), and the environment,
represented by its own short-wavelength 
modes. We compute the decoherence times for the system-field modes and compare 
them with the other time scales of the model. 
\end{abstract}

\maketitle

\section{Introduction}
The emergence of classical physics from quantum behaviour is
important for several physical phenomena in the early Universe.
This is beyond the fundamental requirement that only after the
Planck time can the metric of the Universe be assumed to be
classical. For example, the  inflationary era is assumed to have 
been induced by scalar
inflaton fields, with simple potentials \cite{linde}. Such fields
are typically assumed to have classical behaviour, although in
principle a full quantum description should be used.
The origin of large scale structure in the Universe can be traced 
back to quantum fluctuations that,
after crossing the horizon, were frozen and became classical, stochastic,
inhomogeneities \cite{todosLSS}.

It is generally assumed that several phase transitions have
occurred during the expansion of the Universe \cite{old}. As in
the case for the inflaton fields, the (scalar) order parameter
fields that describe these transitions are described classically.
However, the description of early universe phase transitions from
first principles is intrinsically quantum
mechanical \cite{cormier}. 
As a specific application \cite{Kibble} of the previous point, the
very notion of topological defects (e.g. strings and monopoles)
that characterize the domain structure after a finite-time
transition, and whose presence has consequences for the early
universe, is based on this assumption of classical behaviour for
the order parameter \cite{vilen}, as it distributes itself between
the several degenerate ground states of the ordered system. 

In previous publications, we analysed the emergence of a 
classical order parameter during a second order phase 
transition and the role of decoherence in the 
process of topological defect formation 
\cite{lomplb,deconpb,diana,lomplb2}.

In the present paper our concern is directly related 
with the first point above,
the quantum to classical transition of the inflaton.  
Any approach must take into
account both the quantum nature of the scalar field and the
non-equilibrium aspects of the process \cite{calzettahu95}. 
The problem of the quantum
to classical transition in the context of inflationary models was
first addressed by Guth and Pi \cite{guthpi}. In that work, the
authors used an inverted harmonic oscillator as a toy model to
describe the early time evolution of the inflaton, starting from a
Gaussian quantum state centered on the maximum of the potential.
They subsequently showed that, according to Schr\"odinger's
equation, the initial wave packet maintains its Gaussian shape
(due to the linearity of the model). Since the wave function is
Gaussian, the Wigner function is positive for all times. Moreover,
it peaks on the classical trajectories in phase space as the wave
function spreads. The Wigner function can then be interpreted as a
classical probability distribution for coordinates and momenta,
showing sharp classical correlations at long times. In other
words, the initial Gaussian state becomes highly squeezed and
indistinguishable from a classical stochastic process. In this
sense, one recovers a classical evolution of the inflaton rolling
down the hill.

A similar approach has been used by many authors to describe the
appearance of classical inhomogeneities from quantum fluctuations
in the inflationary era \cite{staro}. Indeed, the Fourier modes of
a massless free field in an expanding universe satisfy the linear
equation
\begin{equation}
\phi_k''+(k^2-\frac{a''}{a})\phi_k = 0.
\end{equation}
For sufficiently long-wavelengths ($k^2\ll a''/a$), this equation
describes an unstable oscillator.
If one considers an initial Gaussian wave function, it will remain
Gaussian for all times, and it will spread with time. As with the
toy model of Guth and Pi, one can show that classical correlations
do appear, and that the Wigner function can again be interpreted
as a classical probability distribution in phase space. (It is
interesting to note that a similar mechanism can be invoked to
explain the origin of a classical, cosmological magnetic field
from amplification of quantum fluctuations).

However, classical correlations are only one aspect of classical
behaviour. It was subsequently recognized that, in order to have a
complete classical limit, the role of the environment is crucial,
since its interaction with the system distinguishes the field
basis as the pointer basis \cite{kiefer}. [We are reminded that,
even for the fundamental problem of the space-time metric becoming
classical, simple arguments based on minisuperspace models suggest
that the classical treatment is only correct because of the
interaction of the metric with other quantum degrees of freedom
\cite{halli}.]

While these linear instabilities cited above characterise   {\it
free} fields, the approach fails when interactions are taken into
account. Indeed, as shown again in simple quantum mechanical
models (e.g. the anharmonic inverted oscillator), an initially
Gaussian wave function becomes non-Gaussian when evolved
numerically with the Schr\"odinger equation. The Wigner function
now develops negative parts, and its interpretation as a classical
probability breaks down \cite{diana}. One can always force the
Gaussianity of the wave function by using a Gaussian variational
wave function as an approximate solution of the Schr\"odinger
equation, but this approximation deviates significantly from the
exact solution as the wave function probes the non-linearities of
the potential \cite{diana,stancioff}.

When interactions are taken into account, classical behaviour is
recovered only for "open systems", in which the unobservable
degrees of freedom interact with their environment. When this
interaction produces {\it both} a diagonalization of the reduced
density matrix and a positive Wigner function, the quantum to
classical transition is completed \cite{giulinibook}.

In Ref.\cite{diana} we have considered an 
anharmonic inverted oscillator coupled to a high temperature
environment. We showed that it becomes classical very quickly, 
even before the wave function probes the non-linearities 
of the potential.  Being an early time event, the quantum to 
classical transition can now
be studied perturbatively. In general, recoherence effects are not
expected \cite{nuno}. Taking these facts into account, we have
extended the approach to field theory models \cite{lomplb,deconpb}. In
field theory, one is usually interested in the long-wavelengths of
the order parameter. Even the early universe is replete with
fields of all sorts which comprise a rich environment, in the 
inflationary example, we considered a model in which the system-field
interacts with the environment-field, including only 
its own short-wavelengths. This is enough during inflation. 
Assuming weak self-coupling constant (inflaton potential is flat) 
we have shown  that decoherence is a short
time event, shorter than the time $t_{\rm end}$, which is
essentially the time by which different modes in the 
system sector cross the horizon. As a result, perturbative
calculations are justified\cite{deconpb}. Subsequent dynamics can
be described by a stochastic Langevin equation, the details of
which are only known for early times \cite{matacz}.

In our approach, the quantum to classical
transition is defined by the diagonalization of the reduced
density matrix. In phase
transitions the separation between long and short-wavelengths is
determined by their stability, which depends on  the parameters of
the potential. During Inflation, this separation is set 
by the existence of the Hubble radius. Modes cross the aparent 
horizon during their evolution, and they are usually treated 
as classical. The main mativation of this talk is to 
present a formal way to understand this statement within 
the open quatum sytem approach. In the last sense, decoherence 
is the critical ingredient if we are to dynamically demonstrate 
the quantum-to-classical transition of the open system.

The paper is organized as follows. In Section 2 we introduce our
model. This is a theory containing a real system field $\phi$, 
massless and minimally coupled to de Sitter background. We compute the
influence functional by integrating out
the environmental sector of the field, composed by the short-wavelength 
modes. Section 3 is
dedicated to reviewing the evaluation of the master equation and
the diffusion coefficients which are relevant in order to study
decoherence. In Section 4 we evaluate upper bounds on the
decoherence times. As we will see, decoherence takes place before the 
end of the inflationary period. Section 5 contains our final remarks.

\section{The Influence functional and the density matrix}

Let us consider a massless quantum scalar field, 
minimally coupled to a de Sitter spacetime  
$ds^2 = a(\eta ) [d\eta^2 - d\vec x^2]$ 
(where $\eta$ is the conformal time $d\eta = dt/a(t)$), 
with a quartic self-interaction. The classical action is given by

\begin{equation}\label{desitteraction}
S[\phi] = \int d^4x ~a^4(\eta )~\left[\frac{{\phi'}^2}{2a^2(\eta )} - 
\frac{{\nabla\phi}^2}{2a^2(\eta )} - \lambda \phi^4\right],
\end{equation} 
where $a(\eta) = -1/(H\eta)$ and $\phi'=d\phi/d\eta$ 
($a(\eta_i) = 1$ [$\eta_i = - H^{-1}$], and 
$H$ is the Hubble radius).
Let us make a system-environment field splitting 

\begin{equation}
\phi = \phi_< + \phi_>,
\end{equation}
where the system field contains the modes with wavelengths longer than the 
critical value $\Lambda^{-1}= 2\pi/\lambda_c$, while the bath field contains 
wavelengths shorter than $\Lambda^{-1}$. As we set $a(\eta_i) = 1$, a physical 
length $\lambda_{\rm phys} = a(\eta)\lambda$ coincides with the corresponding 
comoving length at the initial time. Therefore, the splitting between system and 
environment gives a system sector constituted by all the modes with physical 
wavelengths shorter than the critical length $\lambda_c$ at the initial 
time $\eta_i$.

After splitting, the total action (\ref{desitteraction}) can be written as

\begin{equation}S[\phi] = S_0[\phi_<] + S_0[\phi_>] + S_{\rm int}[\phi_<, 
\phi_>],\label{actions}\end{equation}
where $S_0$ denotes the free field action and the interaction term 
is given by
\begin{eqnarray}&& S_{\rm int}[\phi_<, \phi_>] = - \lambda \int d^4x ~a^4(\eta)~
\left\{\phi_<^4(x) + \phi_>^4(x) \right. \nonumber \\
&+& \left. 6 \phi_<^2(x) \phi_>
^2(x) + 4 \phi_<^3(x) 
\phi_>(x) + 4 \phi_<(x) 
\phi_>^3(x)\right\}.\label{inter}\end{eqnarray}

The total density matrix (for the system and bath fields) is defined
by 
\begin{equation} \rho[\phi_<^+,\phi_>^+,\phi_<^-,\phi_>^-,t]=\langle\phi_<^+ 
\phi_>^+\vert {\hat\rho} \vert \phi_<^- 
\phi_>^-\rangle,\label{matrix}
\end{equation}
where $\vert \phi_<^\pm\rangle$ and $\vert \phi_>^\pm \rangle$ are the 
eigenstates of 
the field operators ${\hat\phi}_<$ and ${\hat\phi}_>$, respectively.
For simplicity, we will assume that 
the interaction is turned on
at the  
initial time $\eta_i$ and that, at this time, the system and the environment are 
not correlated (we ignore, for the moment, the physical consequences of such a 
choice, it has been disussed in \cite{deconpb}). Therefore, the total density matrix can 
be written as the product of 
the density matrix operator for the system and for the bath 
\begin{equation}{\hat\rho}[\eta_i] = {\hat\rho}_{<}[\eta_i] 
{\hat\rho}_{>}[\eta_i].\label{sincorr}\end{equation} We will further assume that 
the initial state of the environment is the vacuum. 

We are interested in the influence of the environment on the evolution of the 
system. Therefore the reduced 
density matrix is the object of relevance. It is defined by 
\begin{equation}\rho_{\rm r}[\phi_<^+,\phi_<^-,\eta] = \int {\cal D}\phi_> 
\rho[\phi_<^+,\phi_>,\phi_<^-,\phi_>,\eta ].\label{red}
\end{equation}
The reduced density matrix evolves in time by means of
\begin{equation}\rho_{\rm r}[\eta ] = \int d\phi_{<i}^+
\int d\phi_{<i}^- ~ J_{\rm r}[\eta , \eta_i]~ 
\rho_{\rm r}[\eta_i],\label{evol}
\end{equation} 
where $J_{\rm r}[\eta,\eta_i]$ is the reduced evolution operator 
\begin{eqnarray}J_{\rm r}[\phi_{<f}^+,\phi_{<f}¯,\eta & \vert & \phi_{<i}^+,\phi_{<i}^-,\eta_i] = 
\int_{\phi_{<i}^+}^{\phi_{<f}}{\cal D}\phi_< \int_{\phi¯_{<i}}^{\phi_{<f}¯}{\cal 
D}\phi_<¯ \nonumber \\
&\times& \exp{\frac{i}{\hbar}\{S[\phi_<^+] - S[\phi_<^-]\}} 
F[\phi_<^+,\phi_<^-].\label{evolred}
\end{eqnarray}
The  influence functional (or Feynman-Vernon functional) 
$F[\phi_<^+,\phi_<^-]$ is defined as 
\begin{eqnarray}
&& F[\phi^+_<,\phi^-_<] = \int d\phi^+_{{> i}} \int d\phi^-_{{>i}} ~ \rho_{\phi_>}
[\phi_{{>i}}^+,\phi_{{>i}}^-,\eta_i] \int
d\phi_{{>f}} \nonumber \\ 
&& \times \int_{\phi^+_{{>i}}}^{\phi_{{>f}}}{\cal D}\phi^+_{>} 
\int_{\phi^-_{{>i}}}^{\phi_{{>f}}}
{\cal D}\phi^-_{>} \exp{\left(i \{S[\phi^+_{>} ]+S_{{\rm int}}
[\phi^+_<,\phi^+_{>} ] \} \right) } \nonumber \\
&&\times  \exp{\left(-i\{S[\phi^-_{>}] + S_{{\rm
int}}[\phi^-_<,\phi^-_{>}]\} \right)}. \nonumber
\end{eqnarray}
This functional takes into account the effect of the 
environment on the system. The influence functional describes the 
averaged effect of the environmental degrees of freedom on the 
system degrees of freedom to which they are coupled. With this 
functional, one can indetify a noise and dissipation kernel 
related by some kind of fluctuation-dissipation relation. This 
relation is important when one is interested in possible stationary 
states where a balance in eventually reached. During inflation we 
have a very flat potential well away from its minimum, and we are, 
in general, only interested in the dynamics over 
some relatively small time. For example, we would neglect 
dissipation during the slow roll period; but it is not 
correct during the eventual reheating phase.

We define the influence action $\delta A[\phi_<^+,\phi_<^-]$ 
and the coarse grained effective action (CGEA) $A[\phi_<^+,\phi_<^-]$ as
\begin{equation}F[\phi_<^+,\phi_<^-] = \exp {\frac{i}{\hbar} 
\delta A[\phi_<^+,\phi_<^-]},\label{IA}\end{equation}
\begin{equation}A[\phi_<^+,\phi_<^-] = S[\phi_<^+] - S[\phi_<^-] + \delta 
A[\phi_<^+,\phi_<^-].\label{CTPEA}\end{equation} 
We will calculate the influence action perturbatively in $\lambda$
and we will consider only terms up to order $\lambda^2$ and one loop in 
the $\hbar$ expansion. The 
influence action has the following form
\begin{eqnarray}
& &\delta A[\phi_<^+,\phi_<^-] = \left\{\langle 
S_{\rm int}[\phi_<^+,\phi_>^+]\rangle_0 - \langle 
S_{\rm int}[\phi_<^-,\phi_>^-]\rangle_0\right\}\nonumber \\
&&+\frac{i}{2}\left\{\langle S_{\rm int}^2[\phi_<^+,\phi_>^+]\rangle_0 - \big[\langle 
S_{\rm int}[\phi_<^+,\phi_>^+]\rangle_0\big]^2\right\}\nonumber \\
&&- i\left\{\langle S_{\rm int}[\phi_<^+,\phi_>^+] 
S_{\rm int}[\phi_<^-,\phi_>^-]\rangle_0\right. \nonumber \\
&&- \left.\langle S_{\rm int}[\phi_<^+,\phi_>^+]\rangle_0\langle 
S_{\rm int}[\phi_<¯,\phi_>¯]\rangle_0\right\} \label{inflac} \\
&&+\frac{i}{2}\left\{S^2_{\rm int}[\phi_<^-,\phi_>^-]\rangle_0 - 
\big[\langle 
S_{\rm int}[\phi_<^-,\phi_>^-]\rangle_0\big]^2\right\},\nonumber
\end{eqnarray} 
where $\langle ~\rangle_0$ is the quantum average, assuming the environment is initially 
in its vacuum state.

The influence functional can be computed, and the result is

\begin{eqnarray}  {\rm Re}\delta A=&& - \lambda \int 
d^4x ~ a^4(\eta)~P(x) \nonumber \\ 
&&+ \lambda^2 \int d^4x\int d^4y ~a^4(\eta)~a^4(\eta') ~ \theta (\eta - \eta') \nonumber \\
&&\times \left\{64 \Delta_3(x) {\rm Re}G^\Lambda_{++}(x-y) 
\Sigma_3(y) \right. \nonumber \\
&&\left. + 288 \Delta_2(x) {\rm Im}G^{\Lambda 2}_{++}(x-y) \Sigma_2(y)\right\},
\label{inff}\\
{\rm Im}\delta A  =&&
- \lambda^2 \int 
d^4x \int d^4y ~a^4(\eta)~ a^4(\eta') \nonumber \\
&& \left\{- 32 \Delta_3(x) {\rm Im}G^\Lambda_{++}(x-y) 
\Delta_3 (y)\right. \nonumber \\ 
&& \left.+ 144 \Delta_2(x) {\rm Re}G^{\Lambda 2}_{++}(x-y) 
\Delta_2 (y) \right\}, \end{eqnarray}
where $G^{\Lambda}_{++}(x-y)$ is the Feynmann propagator 
of the environment field, where the integration over momenta 
is restricted by the presence of the infrared cutoff $\Lambda$. 
We have also defined,

\begin{eqnarray}P = \frac{1}{2}(\phi^{+4}_< 
- \phi^{-4}_<) & ; & \Delta_3 =\frac{1}{2}(\phi_<^{+3} - \phi^{-3}_<)\nonumber \\
\Delta_2 
=\frac{1}{2}(\phi_<^{+2} - \phi^{-2}_<) & ; &\Sigma_3 =\frac{1}{2}(\phi_<^{+3} + 
\phi_<^{-3})\nonumber \\
\Sigma_2 =\frac{1}{2}(\phi_<^{+2} + 
\phi_<^{-2}) & . & \nonumber \end{eqnarray}

\section{Master equation and diffusion coefficients}

In this Section we will obtain the evolution equation for the
reduced density matrix (master equation), paying particular
attention to the diffusion terms, which are responsible for
decoherence. We will closely follow the quantum Brownian motion
(QBM) example \cite{unruh,qbm}, translated into quantum field
theory \cite{lomplb,lombmazz}.

The first step in the evaluation of the
master equation is the calculation of the density matrix
propagator $J_{\rm r}$ from Eq.(\ref{evolred}). In order to solve
the functional integration which defines the reduced propagator,
we perform a saddle point approximation
\begin{equation}
J_{\rm r}[\phi^+_{<f},\phi^-_{<f},\eta\vert\phi^+_{<i},\phi^-_{<i}, \eta_i]
\approx \exp{ i A[\phi^+_{<\rm cl},\phi^-_{<\rm cl}]}, \label{prosadle}
\end{equation}
where $\phi^\pm_{<\rm cl}$ is the solution of the equation of
motion ${\delta Re A/\delta\phi_<^+}\vert_{\phi_<^+=\phi_<^-}=0$
with boundary conditions $\phi^\pm_{<\rm cl}(\eta_i)=\phi^\pm_{<i}$
and $\phi^\pm_{<\rm cl}(\eta)=\phi^\pm_{<f}$. Since we are working up to 
$\lambda^2$ order, we can evaluate the influence 
functional using the solutions of the free field equations.
This classical equation is
$
\phi_<^{''} + 2 {\cal H}\phi_<^{'} - \nabla^2\phi_< = 0,
$ (${\cal H} = a'(\eta)/a(\eta)$). A Fourier mode $\phi_<^{\rm cl}(x) = 
\int_{\vert \vec k\vert < \Lambda} \phi_{\vec k}^{\rm cl}\exp\{i \vec k .\vec x\}$, 
satisfies 

\begin{equation}
\psi_{\vec k}^{''} + \left(k^2 - \frac{a''(\eta)}{a(\eta)}\right)\psi_{\vec k} = 0,
\label{newmodes}\end{equation}
where we have used $\psi_{\vec k} = a(\eta) \phi_{\vec k}$ and the fact that 
$\frac{a''(\eta)}{a(\eta)} = \frac{2}{\eta^2}$. It is important to note that for 
longwavelength modes, $k \ll 2/\eta^2$, Eq. (\ref{newmodes}) describes an unstable 
(upside-down) harmonic oscillator \cite{guthpi}.

As a classical solution can be written as

\begin{equation}
\phi_{\vec k}^{\pm \rm cl}(\eta) = \phi_i^\pm(\vec k)u_1(\eta, \eta_f) + 
\phi_f^\pm(\vec k)u_2(\eta, \eta_f),\label{classmode}\end{equation}
where 

\begin{eqnarray}
u_1 = \frac{\sin[k(\eta - \eta_f)](\frac{1}{k} + k\eta\eta_f)+ 
\cos[k(\eta - \eta_f)] (\eta_f - \eta)}{\rm Idem ~ num. ~ 
\eta \rightarrow \eta_i},\nonumber \\
u_2 = \frac{\sin[k(\eta_i - \eta)](\frac{1}{k} +k\eta\eta_i)+ 
\cos[k(\eta - \eta_i)](\eta - \eta_i)}{\rm Idem ~~ num. ~~ 
\eta \rightarrow \eta_f}.\nonumber 
\end{eqnarray}
We will assume that the system-field 
contains only one Fourier mode with $\vec k = \vec k_0$. This is a sort of 
``minisuperspace" approximation for the system-field that will greatly 
simplify the calculations, therefore we assume 

\begin{equation}
\phi_{<}^{\pm\rm cl}(\vec x, \eta) = 
\phi_{\vec k_0}^{\pm \rm cl}(\eta) \cos({\vec k}_0 .\vec x),
\end{equation}
where $\phi_{\vec k_0}^{\pm \rm cl}$ is given by (\ref{classmode}).

In order to obtain the master equation we must compute the final time 
derivative of the
propagator $J_{\rm r}$.
After that, all the dependence on the initial field configurations 
$\phi^\pm_{<i}$ (coming from
the classical solutions $\phi^{\pm\rm cl}_<$) must be eliminated. In previous 
publicatiuons, 
we have shown that the free propagator satisfies \cite{deconpb}

\begin{eqnarray}
\phi^{\pm\rm cl}_<(\eta) J_0 & = &
\Big[ \phi^\pm_{<f} [u_2(\eta,\eta_f) - \frac{{u'}_2(\eta_f,\eta_f)}
{{u'}_1(\eta_f,\eta_f)}u_1(\eta,\eta_f)] \nonumber \\
& \mp &  i \frac{u_1(\eta,\eta_f)}{{u'}_1(\eta_f,\eta_f)}
\partial_{\phi^\pm_{<f}}\Big]J_0.
\label{rel1}
\end{eqnarray}
These identities allow us to remove the initial field
configurations  $\phi^\pm_{\rm i}$, by expressing them in terms of
the final fields  $\phi^\pm_{\rm f}$ and the derivatives
$\partial_{\phi^\pm_{\rm f}}$, and obtain the master equation.

The full equation is very complicated and, as for quantum Brownian
motion, it depends on the system-environment coupling. In what
follows we will compute the diffusion coefficients for the
different couplings described in the previous section. As we are
solely interested in decoherence, it is sufficient to calculate
the correction to the usual unitary evolution coming from the
noise kernels (imaginary part of the influence action). The result
reads
\begin{eqnarray}&&i \hbar \partial_\eta \rho_{\rm r}[\phi_{<f}^+,\phi_{<f}^-,\eta]= 
\langle \phi_{<f}^+\vert\Big[{\hat H}_{\rm ren}, 
{\hat\rho}_{\rm r}\Big]\vert\phi_{<f}^-\rangle \nonumber \\
&-& i \left[\Gamma_1 ~D_1({\vec k}_0;\eta)
+ \Gamma_2 ~ D_2({\vec k}_0;\eta) \right] \rho_r[\phi^+_{<f},\phi^-_{<f},t] \nonumber \\
&+&  ...~ ,\label{master}\end{eqnarray} 
where we have defined $\Gamma_1 = \frac{V \lambda^2}{\Lambda^{-3}}\frac{(\phi_{<f}^{+3} 
- \phi_{<f}^{-3})^2}{H^5}$ and $\Gamma_2 = \frac{V\lambda^2}{\Lambda^{-3}}\frac{
(\phi_{<f}^{+2} - 
\phi_{<f}^{-2})^2}{H^3}$. $V$ is the spatial volume inside 
which there are no coherent superpositions of macroscopically distinguishable 
states for the system field. The ellipsis denotes other terms 
coming from the time derivative that not contribute to the diffusive effects. 
This equation contains time-dependent 
diffusion coefficients $D_i(t)$. Up to one loop, only $D_1$ and $D_2$ survive.
Coefficient $D_1$ is related to the interaction term $\phi_<^3 \phi_>$, while 
$D_2$ to $\phi_<^2 \phi_>^2$. These coefficients can be (formally) written as

\begin{eqnarray}
D_1(k_0,\eta) &=& 2\frac{H^5}{\Lambda^3}\int_{\eta_i}^{\eta} d\eta'
a^4(\eta) a^4(\eta') F_{\rm cl}^3(k_0,\eta,\eta') \nonumber \\
&\times& {\rm Im}G_{++}^{\Lambda}(3k_0,\eta, \eta')
~ \theta (3k_0 - \Lambda),
\end{eqnarray}
and 
\begin{eqnarray}
&&D_2(k_0,\eta) = 36 \frac{H^3}{\Lambda^3}
\int_{\eta_i}^{\eta}d\eta' a^4(\eta) a^4(\eta') F_{\rm cl}^2(k_0,\eta,\eta') 
\nonumber \\
&\times& \left[{\rm Re}G_{++}^{\Lambda 2}(2k_0,\eta, \eta') + 
2{\rm Re}G_{++}^{\Lambda 2}(0,\eta, \eta')\right] ,
\end{eqnarray}
where the function $F_{\rm cl}$ is

\begin{equation}
F_{\rm cl}(k_0,\eta,\eta') = \frac{\sin[k_0(\eta - \eta')]}{k_0\eta} + 
\frac{\eta'}{\eta} \cos[k_0(\eta - \eta')].\end{equation}

The explicit expression of these coefficient are complicated 
functions of conformal time, the particular mode $k_0$, and the 
cutoff $\Lambda$, and we  will  show them 
in a separate publication \cite{dianaln}. For the porpuse of this talk, 
we will use some analytical approximations, which will allow 
to obtain an estimation 
of the scale of decoherence in a particular case.

\section{Decoherence}

The effect of the diffusion coefficient on the decoherence process
can be seen considering the following approximate solution to the
master equation
\begin{eqnarray} &\rho_{\rm r}&[\phi^+_<, \phi^-_<; \eta] \approx 
\rho^{\rm u}_{\rm r}[\phi^+_<, \phi^-_<; \eta] \nonumber \\
&&\times  \exp \left[-\sum_j \Gamma_{\rm j} \int_{\eta_i}^{\eta_f} 
d\eta ~D_{\rm j}(k_0,\Lambda,\eta)
\right], \end{eqnarray} where $\rho^{\rm u}_{\rm r}$ is the
solution of the unitary part of the master equation (i.e. without
environment), and $\Gamma_j$ includes the 
coefficients in front each diffusion term in Eq.(\ref{master}). 
The system will decohere when the non-diagonal
elements of the reduced density matrix are much smaller than the
diagonal ones.

The decoherence time-scale sets the time after which we have 
a classical field configuration, and it can be defined as the solution 
to 

\begin{equation}
1 \approx \sum_j\Gamma_j\int_{\eta_i}^{\eta_D}d\eta D_j(k_0,\Lambda,\eta ).
\end{equation}
We will solve this equation only in a very approximated way in order 
to find upper bounds for the decoherence times coming from each 
diffusion coefficients. A more refined 
evaluation of decoherence times will be shown in Ref. \cite{dianaln}. For example, 
when we consider a mode $k_0 \leq H$, we can probe that a good approximation 
to $D_1$ is given by

\begin{equation}
D_1^{\rm approx}(k_0 \sim H,\eta) \sim  -\frac{(1 + H\eta)}{H\eta^7k_0^3\Lambda^3}.
\end{equation}
This is valid for modes shorter or the same order than $H$; but it is an overestimation 
for $k_0 \ll H$. As 
$D_1$ is defined under the constraint $\Lambda/3 < k_0 < \Lambda$, and considering 
we will set the critical length $\Lambda \leq H$, ours is a good estimation.
Analysing the coefficient $D_2$, it is possible to find an approximated 
expression  for small values of $k_0$ (respect to $H$). 
For $k_0/H > 1$, that corresponds to those modes out 
of the aparent horizon at $\eta_i$, the diffusion coefficient $D_2$ is an oscilatory 
function and it has a maximun when $k_0 \sim \Lambda$, this is the behavior noted 
for conformally coupled fields \cite{lombmazz}. Finally, for long-wavelength modes, 
we can write

\begin{equation}
D_2^{\rm approx}(k_0 < H,\eta) \sim  \frac{1}{H\eta^4\Lambda^3}.
\end{equation}

Both approximations above, are close to the 
exact coefficients when $\Lambda \leq H$.

In order to quantify decoherence time we have to fix the values of $\Gamma_{1,2}$. For this, 
we have to assume values to $\lambda$, $V$, $\Delta$, and $\Sigma$. We will use the 
more conservative choice in order to have a lower bound to the decoherence time. 

Assuming slow roll condition $1/2 (d\phi/dt) \ll U = \lambda \phi^4$, the classical 
equations (using $\ddot\phi\ll U'$) are

\begin{equation}H^2 = \frac{8\pi U}{3m_{\rm pl}^2} ~~; ~~ \dot\phi = -\frac{U'}{3H},
\end{equation}
these equations are obtained under the following conditions
\begin{equation}\epsilon_U = \frac{m_{\rm pl}^2}{2}\frac{U'}{U}\ll 1 
~~; ~~ \sigma_U = m_{\rm pl}^2 \frac{U''}{U}\ll 1,
\end{equation}
where $m_{\rm pl}^2= 1/G$ is the Plank mass.

Definig the end of the inflationary period setting $\epsilon_U \sim 1$, one can set 
$\phi(N) \approx \sqrt{8N} m_{\rm pl}$, where $N = \ln{a(\eta_f)/a(\eta)}$ is the 
e-fold number. Thus, we assume the mean value of the system field at time of 
decoherence is $\phi$, and we set $(\phi_<^+ - \phi_<^-)\sim 10^{-5}\phi$, and $V \sim H^{-3}$.  

From previous considerations, we can show 

\begin{equation}
t_{D_1} \leq \frac{1}{6H} \ln\left\{\frac{5k_0^3H^3}{\lambda^2 10^{-15}\phi^6}
\right\} \sim  
\frac{1}{6H}\ln\left\{\frac{\lambda \phi^6 10^{15}}{m_{\rm pl}^6}\right\} , 
\end{equation}

and 

\begin{equation}
t_{D_2} \leq \frac{1}{3H} \ln\left\{\frac{3H^4}{\lambda^2 10^{-10}\phi^4}
\right\} \sim \frac{1}{3H} \ln\left\{\frac{10^{10}\phi^4}{m_{\rm pl}^4}
\right\}.\end{equation}

Using  $N = H t_{\rm end} \geq 60$ as an estimative scale to the end of 
inflationary period, 

\begin{equation}\frac{t_{D_1}}{t_{\rm end}}\leq \ln\{\lambda 10^{15}\} \leq 1,
\end{equation}
corresponding to the diffusion term $D_1$ and values of $\lambda \sim 10^{-8}$. 
The scale coming from $D_2$ is

\begin{equation}\frac{t_{D_2}}{t_{\rm end}} \leq \frac{1}{10},
\end{equation}.

From scales $t_{D_1}$, and $t_{D_2}$, we can see that for a given 
mode $k_0 < \Lambda \leq H$, decoherence 
is effective by the time in which inflation 
is ending.

\section{Final remarks}

Let us summarize the results contained in this paper.
After the integration of the high frequency modes in Section 2,  
we obtained the CGEA for the 
low energy modes. From the imaginary part of the CGEA
we obtained, in Section 3, the diffusion coefficients
of the master equation. System and environment are two sectors of a single
scalar field, and the results depend on the  ``size" of
these sectors, which is fixed by
the critical wavelength $\Lambda^{-1}$. 

In Section 4 we analysed the decoherence times for those modes
in the system whose wavelength is shorter than the critical value.

We have shown some analitycal approximations that allow to 
conclude that if we consider a critical length $\Lambda \sim  H$, 
those modes with wavelength $k_0 \ll \Lambda$ are the more 
affected by diffusion throught coefficient $D_2$. For these modes, 
we can show that the effect is not dependent of the critical $\Lambda$ 
\cite{dianaln}.

If one consider a cutoff $\Lambda \geq H$, and modes $H < k_0 < \Lambda$, 
diffusive effects are larger for those modes in the system 
whose wavelength is close to the critical $\Lambda^{-1}$ \cite{lombmazz}.

In Ref. \cite{dianaln} we will present the complete expression 
of the diffusive terms, an also an extensive analysis of the 
evaluation of the timescale for decoherence in several cases of 
interest.

\section{Acknowledgments}
This work is supported by UBA, CONICET and Fundaci\'on Antorchas; Argentina.
It is a pleasure to thank of the organisers, specially to 
H. Thomas Elze, for the invitation and hospitality during the meeting DICE '04 
at Piombino. I also appreciate very interesting conversations with B.L. Hu, C. Kiefer, 
N. Mavromatos, and R. Rivers.

\bibliography{apssamp}

\end{document}